# New Dimension Value Introduction for In-Memory What-If Analysis

GAURAV SAXENA, Times Internet Limited
RUCHI NARULA
MANISH MISHRA, Times Internet Limited

OLAP systems operate on historical data and provide answers to analyst's queries. Philosophy and techniques of what-if analysis on data warehouse and in-memory data store based OLAP systems have been covered in great detail before but exploration of new dimension value (attribute) introduction has been limited in the context of what-if analysis. We extend the approach of Andrey Balmin et al of using select modify operator on data graph to introduce new values for dimensions and measures in a read-only in-memory data stored as scenarios. Our system constructs scenarios without materializing the rows and stores the row information as queries. The rows associated with the scenarios are constructed as and when required by an ad-hoc query.

## 1. INTRODUCTION

OLAP systems have become an indispensable tool for data analysis, trend spotting, decision support and knowledge discovery for executives across industries. These tools help executives observe past trends of data and anticipate the changes in future. A critical part of this process is What-if analysis. What-if analysis can be described as a data intensive simulation whose goal is to inspect the behavior of a complex system (i.e., the enterprise business or a part of it) under some given hypotheses (called scenarios) [Golfarelli et. al., 2006].

Typically, an analyst formulates a hypothetical scenario by changing some attribute of the data which he uses in conjugation with other scenarios and historical data (henceforth referred as real data) to come up with a future strategy. For example, consider a manager in manufacturing unit who needs to ascertain the demand for the year 2012, using the data of 2011. He needs to generate a sourcing plan and the corresponding budget. His sourcing plan for 2011 is shown in the table below

Table I. Sourcing plan of manager for 2011

| Year | Supplier | Product | Volume (MM Tons) | Cost ($/Ton) | Amount (MM$) = Cost * Volume |
|------|----------|---------|------------------|--------------|------------------------------|
| 2011 | SU1 | P1 | 10 | 1.0 | 10 |
| 2011 | SU1 | P2 | 11 | 1.5 | 16.5 |
| 2011 | SU2 | P1 | 12 | 1.1 | 13.2 |
| 2011 | SU2 | P2 | 13 | 1.4 | 18.2 |
| | | | | Total(MM$) | 57.9 |

His question "How would my total amount change if I increase volume of products by 10% for the year 2012?" could be answered by making a scenario for 2012 with 10% increased volume. Typically, OLAP systems do not allow any change in their data because they use immutable data warehouses. A data warehouse is a non-volatile, time-variant and coarse grained collection of data from diverse sources. It is designed for fast data extraction and does not support real-time updates to maintain performance [Elmasri and Navathe, 2006]. It typically uses a multidimensional data model which consists of dimensions and facts (measures) arranged in a star schema to swiftly answer ad-hoc queries. A query is any meaningful combinations of values for the attributes in dimension (dimension value) and fact tables. Traditional OLAP systems could only answer pre-configured queries because they relied on materialized views in a data warehouse. What-if analysis on such systems would mean insertion of new data in the cube and re-computation of the cube. This would need an experienced data warehouse person and a long downtime of the analytical system [Balmin et. al., 2000].

This challenge can be addressed by using multi-version data warehouses which keep track of data changes [Bebel et. al., 2004]. However, this approach is space intensive and may not be suitable for in-memory systems. This is especially true when a large chunk of data is changed, in spite of data sharing [Bebel et. al., 2004] between different versions. To address this problem, a system described in [Zhou et. al., 2009] performs what-if analysis on OLAP cube using query re-writing alleviating the need to maintain different versions of data. Under this system, updates to the cube are not applied and are stored as rules. Any query on the cube is re-written to return data as if it were changed because of the updates. However, this system only allows change in measure values and doesn't allow addition of new dimension values. It also requires scan of all the rows, for all the update rules irrespective of their application to a query.

RDBMS can also be used to mitigate the problem of downtime. New rows of any scenario could be added to the tables in RDBMS in real-time and queries can be served using both the real rows and hypothetical rows. A scheme to implement hypothetical databases using difference lists is described in [Woodfill and Stonebraker, 1983; Stonebraker et. al., 1981]. However, this increases space requirements linearly [Woodfill and Stonebraker; 1983, Stonebraker et. al., 1981] and is inefficient as compared to query re-writing [Zhou et. al., 2009]. Also, with RDBMS, we lose the performance advantage of fast, multi-view, multi-aggregated data cube architecture.

In-memory data stores solve both the problem of a long downtime and fast response time by doing away with pre-aggregations [Schaffner et. al., 2009]. They store compressed data in-memory and can run any ad-hoc query through the data to find its result. Scenario rows can be added to such an in-memory data stores in real time. Since aggregations and views are not pre-computed, response to a query can contain data from both scenario and real rows. This approach is also space intensive, especially, when a large fraction of the cube is changed. In addition, all the above approaches assume that all scenarios are visible to all users and no scenario is private to the analyst making it.

Multiple approaches to keep deltas of changed measures in user space have been reported [Pasumansky and Netz., 2004; Xiao et. al., 2009; Walker et. al.; 2010]. According to these approaches, only the rows which are changed in the data cube are stored as a scenario in the user space. This allows scenario data to be selectively deleted or updated for each user. The real data cube, hence, stays immutable. The change in measure values are stored in scenarios and applied to real values when queried. Space requirement for these approaches is similar to those maintaining difference lists. Moreover, if changes are kept local to a user, space requirements may further increase with the number of users. In our example, if the manager wants to increase the volume of all the products by 10%, the whole data cube needs to be stored in user space with the changed volume. The changed data cube would look like as shown in Table II.

Table II. Sample table for storing deltas in volume

| Year | Supplier | Product | ΔVolume (MM Tons) |
|---|---|---|---|
| 2011 | SU1 | P1 | 1 |
| 2011 | SU1 | P2 | 1.2 |
| 2011 | SU2 | P1 | 1.3 |
| 2011 | SU2 | P2 | 1.4 |

In this paper we describe a way to enable in-memory data stores to introduce new dimension values and store them as scenarios (in user or global space). These scenarios and their corresponding measure value changes can be addressed using the new dimension values in ad-hoc queries. Consequently, the new dimension values can also be a part of filters in an analytics system. We present a system which stores queries for rows inserted due to creation of scenarios. In our example, the manager wants to increase the volume of all the products by 10% for the year 2012. Our system would create a new dimension value 2012 as a scenario and instead of storing the materialized cube shown in Table III, would store a query (select all the rows) and corresponding change associated with the measures as a factor (1.1 times the previous value).

Table III. Data cube as required by manager for 2012

| Year | Supplier | Product | Volume (MM Tons) | Cost ($/Unit) |
|---|---|---|---|---|
| 2012 | SU1 | P1 | 11 | 1.0 |
| 2012 | SU1 | P2 | 12.1 | 1.5 |
| 2012 | SU2 | P1 | 13.2 | 1.1 |
| 2012 | SU2 | P2 | 14.3 | 1.4 |

## 1.1 Contributions

What-if analysis for a data cube using query re-writing has been covered in great detail [Balmin et. al., 2000]. We extend their work for in-memory what-if analysis using a query system described in [Li et. al., 2004] to introduce new dimension values. We present a system of operators which can introduce new values for both dimensions and measures and store them as scenarios without changing the real data cube. This enables insertion of rows with a new dimension value and allows their update and partial or complete deletion. The system imposes no restriction of its use in user or global scope.

We do not materialize the new rows that may be added to the data cube as a result of creation of scenarios. Instead, we store these new rows as a set of queries in scenario definitions. When a query is evaluated on the system, the queries stored as a part of scenarios are used to simulate the new rows. The result containing the simulated rows and the real rows can either be materialized or passed to an accumulator. We present algorithms to execute the above approach.

## 2. FRAMEWORK
### 2.1 Data cube and query

Let us consider an in-memory data store D, henceforth called a data cube, containing measures S and dimensions V. Let $V_i$ be a dimension and let $v_j^i$ be $j^{th}$ dimension value (attribute) of dimension $V_i$. Let there be n dimensions and each dimension $V_i$ contains $n_i$ dimension values. Let each measure be characterized as $S_k$ and measures values in each row are characterized as $s_l^k$ where l is the row number. Measure values are numeric only.

A query Q can be written as $<a_1, a_2, a_3...>$ where $a_i \, \varepsilon \, \{*, v_1^1, v_2^1, ...v_{n_1}^1, v_1^2, ...v_{n_n}^n\}$. * for a dimension $V_k$ includes all its dimension values [Li et. al., 2004]. It is possible to have no dimension values for a dimension in the query. Such a query when run on D returns an empty set.

Each row in a data cube contains at most one dimension value for each dimension. Such a row can be characterized by dimension values which qualify it. If t is a row, then it can be written as
$t \equiv <v_{j_1}^1, v_{j_2}^2, v_{j_3}^3...v_{j_n}^n; S_1: s_{t_l}^1, S_2: s_{t_l}^2...>$
Where $j_k \, \varepsilon \, (1, 2... n_k)$ for a dimension $V_k$ present in t, $\forall \, k <= n$
$t_l$ is the row number.

Let us also break this into a shorter form with the set of dimension value of row be represented as a query $t_d \equiv <v_{j_1}^1, v_{j_2}^2, v_{j_3}^3...>$ and measures be a set $t_s \equiv \{s_t^1, s_t^2...\}$

Let us say the Table IV makes a data cube:

Table IV. Data Cube for 2011

| Year | Supplier | Product | Volume (MM Tons) | Cost ($/Unit) | Amount(MM$) = Cost * Volume |
|---|---|---|---|---|---|
| 2011 | SU1 | P1 | 10 | 1.0 | 10 |
| 2011 | SU1 | P2 | 11 | 1.5 | 16.5 |
| 2011 | SU2 | P1 | 12 | 1.1 | 13.2 |
| 2011 | SU2 | P2 | 13 | 1.4 | 18.2 |

Here Supplier, Product and Year are dimensions and Cost and Volume are measures. Each row in the cube can be written using the query system described above e.g. first row can be written as <SU1, P1, 2011; Volume: 10, Cost: 1.0> where $v_{j_1}^1$= SU1, $v_{j_2}^2$= P2 and $v_{j_3}^3$= 2011, $S_1$= Volume, $S_2$ = Cost $s_1^1$= 10, $s_1^2$= 1.0. Similarly, other rows can be written as follows
Row 2 <SU1, P2, 2011; Volume: 11, Cost: 1.5>
Row 3 <SU2, P1, 2011; Volume: 12, Cost: 1.1>
Row 4 <SU2, P2, 2011; Volume: 13, Cost: 1.4>

If a query Q <*, P1, *> is applied on the data then 1st and 3rd rows form the result set. Here * is used for dimensions supplier and year. Similarly, if another query Q1 <SU1, *, 2011> is applied on the data then 1st and 2nd rows form the result set. Here * is used for product.

### 2.2 Scenario

To define scenarios using our framework, let us consider that we have a set of rows in data cube and each dimension value can define a query to get a subset of rows in which it appears. These rows can be said to be associated with that particular dimension value. Thus, to get a set of rows associated with a dimension value $v_j^i$ the query will be of the form <*,…$v_j^i$, *…> where $v_j^i$ is the only dimension value of dimension $V_i$ present in the query and for all other dimensions * is present e.g. dimension value SU1 is present in row 1 and row 2 of data cube shows in Table IV. The query which fetches rows associated with SU1 can be written as <SU1, *, *>.

To create a scenario, we choose a new dimension value, select a sub-cube using queries, change its measure values and substitute the 'scenario' dimension value in the sub-cube. Just like a real dimension value, we can also say that these rows are associated with the scenario dimension value. Instead of storing the materialized sub-cube, we store the corresponding queries.

Let us represent the set of scenarios by W. A scenario $W_k$ in W contains scenario dimension value $w_k$ created for a dimension $V_{w_k}$. Let us also consider a collection of queries $Q_{w_k}$ which represents rows $w_k$ is associated with. Since there cannot be a real row t which can have a $t_d$ containing $w_k$, we propose that $Q_{w_k}$ fetch real rows only. We also propose that a scenario $W_k$ stores mapping from the dimension $V_{w_k}$ to the scenario dimension value $w_k$. In addition, each query in $Q_{w_k}$ contain a factor for each measure by which respective measure values in the result set of the query is to be multiplied. To materialize the rows associated with $w_k$, result set of $Q_{w_k}$ is obtained from the data cube D, all dimension values of $V_{w_k}$ are replaced by $w_k$ and measure values of rows are multiplied by the respective factors. We will explain all these points below using our example. We also propose that the queries of a scenario and factors associated with each query can be edited. However, the mapping from dimension to new dimension value is immutable and stays the same for all queries. At this point we also define R to be fictitious materialized data cube containing rows of all the real and scenario dimension values. A scenario W is written as
$< w, V_i: b_1 >$
Where
$V_i: b_1$ denotes that all dimension values of a dimension $V_i$ in the result set will be replaced by a new dimension value $b_1$
$w$ is the scenario name.

A query in $Q_{w_k}$ for the scenario can be written as $< a_1, a_2, a_3 \ldots; s_1: f_1, s_2: f_2 \ldots >$
Where $a_i \varepsilon \{*, v_1^1, v_2^1, \ldots v_{n_1}^1, v_1^2, \ldots v_{n_n}^n \}$, * for a dimension $V_i$ includes all its dimension values
$s_1: f_1, s_2: f_2 \ldots$ denotes that in the result set, measure values of measure $s_j$ are to be multiplied by $f_j$.

In our example, sourcing manager needs to compute the volume requirements for year 2012 to make his projections. Thus, he makes a scenario for 2012 $< 2012, Year: 2012 >$. According to the expression of scenario, $w = 2012, V_i = Year, b_i = 2012$. He then associates the rows of data cube (shown in Table IV) with the scenario using the following queries. These queries form the collection $Q_{2012}$
q1 <SU1, *, 2011; Volume: 2, Cost: 1> 1st and 2nd rows qualify for this query
q2 <SU2, *, 2011; Volume: 3, Cost: 1> 3rd and 4th rows qualify for this query
Substituting the scenario dimension value 2012 for year in the result set of these queries we get the rows in Table V for $W_{2012}$. These rows are shown for illustration only and are not materialized

Table V. Data Cube for $W_{2012}$ with 2012 substituted

| Query | Year | Supplier | Product | Volume (MM Tons) | Cost ($/Unit) | Amount(MM$) = Cost * Volume |
|---|---|---|---|---|---|---|
| q1 | 2012 | SU1 | P1 | 10 | 1.0 | 10 |
|    | 2012 | SU1 | P2 | 11 | 1.5 | 16.5 |
| q2 | 2012 | SU2 | P1 | 12 | 1.1 | 13.2 |
|    | 2012 | SU2 | P2 | 13 | 1.4 | 18.2 |

When the factors are multiplied with the measure values of the row, the result set becomes as shown in table VI.

Table VI. Data Cube for $W_{2012}$ with measure values multiplied by factors

| Query | Year | Supplier | Product | Volume (MM Tons) | Cost ($/Unit) | Amount(MM$) = Cost * Volume |
|---|---|---|---|---|---|---|
| q1 | 2012 | SU1 | P1 | 20 | 1.0 | 20 |
|    | 2012 | SU1 | P2 | 22 | 1.5 | 33 |
| q2 | 2012 | SU2 | P1 | 36 | 1.1 | 39.6 |
|    | 2012 | SU2 | P2 | 39 | 1.4 | 54.6 |

### 3. OPERATORS

#### 3.1 Operators on the data cube

Addition operator + takes as operands two data cubes with identical dimensions and measures. It returns a new data cube containing all the records of the two data cubes. Same records in the new data cube can occur multiple times. The expression $\sum_i X_i$ is used to represent a series of + operators acting on multiple operands i.e. $\sum_i X_i = X_1 + X_2 + X_3 + \ldots$

Select operation σ [Balmin et. al., 2000] takes two operands: a data cube D and a query q. It returns a subset of D containing all rows t of D such that $t_d$ is a subset of q i.e. $\sigma_{D,q} \equiv \{t \mid t \in D, t_d \subset q\}$

1. σ returns an empty system Φ if acted on an empty system Φ
2. σ returns D if q is <*… *>
3. σ returns Φ if ∀ rows t of D, $t_d \not\subset q$

We extend select modify operator [Balmin et. al., 2000] $\sigma^{\wedge}$ to take as operands a mapping m from a dimension to a new dimension value, in addition to a data cube D and a set of multiplying factors e, one for each measure. Also, in our system, unlike Balmin et.al. [2000], select modify operator doesn't take query as a parameter. It returns a new system which has the same number of records as D but with dimension and measure values replaced with new ones respectively

1. $\sigma^{\wedge}_{D,e,m}$ returns an empty set Φ if D is empty
2. $\sigma^{\wedge}_{D,e,m}$ returns D if both e and m are empty

#### 3.2 Operators on the queries

Scenario extractor operation π takes as operand a query Q and returns a set of scenarios for which dimension values are present in Q i.e. $\pi_Q \equiv \{W_k \mid W_k \in W, w_k \in Q\}$

Real sub query operator ρ when applied on query Q removes all scenario dimension values and returns a query with only real dimension values of Q. When no real dimension values are present for a dimension in the query, * is put for the dimension.

Query resolution operator μ when applied on sets of queries does the following

1. It takes Cartesian product of the first two sets of queries. The resulting element in the Cartesian product consists of two queries, one from each set. These queries are intersected with each other. This process is repeated for each element of the Cartesian product.

2. If it is acting on multiple sets of queries, the resulting set of queries from the first operation is treated as a new set and the same procedure is applied until only one set is left
i.e. $\mu_{q_1,q_2…q_n} = \mu_{q_1 \times q_2, q_3…q_n} = \mu_{q_{12}, q_3…q_n} … = \mu_{q_{12…n}}$ (1)
Where
$q_1$, $q_2$ are sets of queries
$\mu_{q_{12…n}}$ is the final set containing all the resulting queries
× represents Cartesian product
$q_{12}$ is the set of queries formed by intersection of each element of the set formed by taking Cartesian product of $q_1$ and $q_2$. If two queries, being intersected, have complimentary dimension values for a dimension $V_i$, their intersection contains no dimension values for $V_i$. Such a query is removed from the set as it cannot fetch any row from data cube. As a result, if result set of any intermediate or final query (like $q_{12}$) is Φ, μ also returns Φ
3. When $q_i$ and $q_j$ with multiplying factors $e_i$ and $e_j$ respectively are intersected to give a query $q_{ij}$ its multiplying factor $e_{ij} = e_i * e_j$
4. It returns the resulting set of queries. In case a single set is passed to it, the same is returned without modification

Query combination operator ɳ takes a query as operand and returns a set of queries with all the combinations of the scenario dimension values present in a query such that

1. There is at least one dimension value (real or scenario) for each dimension
2. If a scenario dimension value is present for a dimension, it cannot have any other dimension values – real or scenario.
3. Real dimension values of a dimension always occur together and are not further divided

Such queries are called atomic queries. If any of the above conditions are not met for any query, then it returns Φ.

### 3.3 Operators on scenarios
Scenario Queries operator θ acts on a set of scenarios $W'$ and returns a map M with a query as key and collection of queries as its value i.e. $\theta_{W'} \equiv M \equiv \{K: K_{W_i}^j, V: \{Q_{w_{i_1}}^j, Q_{w_{i_2}}^j …\} \mid \forall\, K_{W_i}^j \in K_{w_i}, \forall\, W_i \in W'\}$
where
K stands for key and V for value
$K_{W_i}$ is the set of keys for scenario $W_i$

### 4. SYSTEM DEFINITION
### 4.1 Scenario definition
A sub-cube $D_1$ of scenario $W_1$ can be written using our operators as
$$D_1 = \sum_i \sigma_{\hat{\sigma}_{D,q_i}; S_1:e_i^1, S_2:e_i^2 …; V_j: w_1}$$ (2)
Where
$W_1$ is a scenario made on data cube D
$q_i$ is a query corresponding to rows in $D_1$. The set $\{q_1, q_2…\} \equiv Q_{w_1}$
$e_j^i$ is a factor against query $q_i$ and measure $S_j$
$w_1$ is the scenario dimension value which replaces all dimension value of dimension $V_j$

The 2012< 2012, Year: 2012 > scenario of sourcing manager can be written with the help of our operators as follows. The result of this operation is shown in Table VII and is identical to Table VI

$$< 2012, \text{Year}: 2012 > \equiv \sigma^{\wedge}_{\sigma_{D,<SU1,*,2011>}; \text{Volume 2,Cost 1;Year:2012}} +$$

$$\sigma^{\wedge}_{\sigma_{D,<SU2,*,2011>}; \text{Volume 3,Cost 1;Year:2012}} \quad (3)$$

Table VII. Sub cube of scenario $W_{2012}$

| Year | Supplier | Product | Volume(MM Tons) | Cost($/Unit) | Amount(MM$) = Cost * Volume |
|------|----------|---------|-----------------|--------------|------------------------------|
| 2012 | SU1 | P1 | 20 | 1.0 | 20 |
| 2012 | SU1 | P2 | 22 | 1.5 | 33 |
| 2012 | SU2 | P1 | 36 | 1.1 | 39.6 |
| 2012 | SU2 | P2 | 39 | 1.4 | 54.6 |

As shown in the example, we can create scenarios from a cube D. These scenarios define their own cubes viz. $D_1$, $D_2$, $D_3$... which can be visualized as nodes of a directed data graph [Balmin et. al., 2000]. We can now answer the following questions on this graph

1. Any two nodes in the data graph can be compared for variance
2. Data of any number of nodes can be aggregated according to a query
3. Changes in the data graph due to change of dimensions or measures in a particular node can be evaluated. This would need simultaneous evaluation of many nodes of data graphs at the same time [Balmin et. al., 2000]

### 4.2  Scenario Implementation

Let us say that rows of a new scenario dimension value will be corresponding to a query Q. In our system, Q can only be defined in terms of real dimension values (henceforth called real query)
$Q \equiv < a_1, a_2, a_3... >$ where $a_i \; \varepsilon \; (*, v_1^1...v_{n_n}^n)$ (4)
Let us say that the set of $u_k$ queries $Q_{w_k} \equiv \{Q_1^k, Q_2^k...Q_{u_k}^k\}$ is associated with a scenario $W_k$. If a new query $Q_v^k$ containing any combination of real and scenario dimension values (henceforth called arbitrary query) is associated with a scenario, it is reduced to real queries. The query $Q_v^k$, then, is stored in a two tier structure: an atomic query as key (hereafter called key query) and a collection of real queries against it. This is done in two steps:

1. $Q_v^k$ is broken into key queries using operator η
2. Each key query is reduced to real queries using operator μ

Such a data structure helps in evaluation of an arbitrary query over the fictitious materialized data cube R containing rows of all dimension values, real or scenario. This will be covered in detail later in the section 4.5.

A real row t and a real query Q, represent a row of $w_k$, if $t_d \subset Q$ for some v, $0 < v \leq u_k$. Measure values $t_s$ of such a row are captured by multiplicative factors $\{e^k_{v\,S_1}, e^k_{v\,S_2}...\}$ where $e^k_{v\,S_i}$ is a factor corresponding to measure $S_i$. The measure value of this row is computed by multiplying the factor with the measure value of the real row t i.e. $t_{s_1} e^k_{v\,S_1}$ for $S_1$, $t_{s_2} e^k_{v\,S_2}$ for $S_2$ etc.

A real query Q associated with a scenario and an eligible real row t for the query, together, can give rise to only one simulated row because a row can be found eligible in a query only once. Though, the same row may be found eligible with another query and may be used to make another simulated row. Similarly, more than one row can be found eligible for a query.

### 4.3  Dependent scenarios
Scenarios when made using the result set of another scenario make the former dependent on the latter. This dependence is not limited to two scenarios only and can be of any width, i.e. a scenario can depend on any number of scenarios and any depth, i.e. a scenario can depend on any number of scenarios which in turn can depend on any number of scenarios and so on. This results in a directed graph of scenario nodes, each representing a result set corresponding to it.

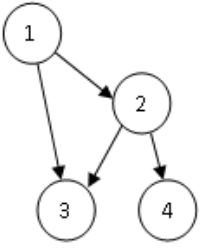

Fig. 1. Data Graph

This is similar to the data graph discussed in [Balmin et. al., 2000]. In the data graph shown in Fig 1, node 1 is the node containing real rows D. Node 2 is a scenario $W_2$ made on a subset of D. Node 3 was then made as a union of a subset of D and a subset of $W_2$. Node 4 was made only with a subset of $W_2$. For an arbitrary query, the result set will be a bag of rows drawn from each node by running the query through its collection of rows. This operation would require complete graph traversal for each scenario in the query and thus would mean exponential execution time [Balmin et. al., 2000].

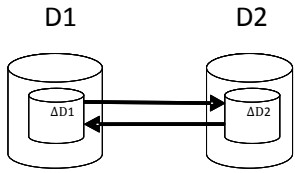

Fig 2. Dependent Scenarios

Having a structure like this also means that each scenario will be affected by the changes in the scenario it is dependent on directly and indirectly. A few nodes can even form cycles among themselves. These cyclic dependencies can result in an infinite loop when queries are executed. Cyclic dependencies between various scenarios may not necessarily cover all the rows e.g. in Fig 2, $D_1$ and $D_2$ are nodes of a data graph. $D_1$ was created as a result of query q and $q_1$ on D and it introduced dimension value $c_1'$ for dimension $V_1$. Similarly $D_2$ was created as a result of query $q_2$ on $D_1$ and $q_3$ on D and it introduced dimension value $c_1''$ for dimension $V_1$

$$D_1 = \hat{\sigma}_{\sigma_{D,q};S_1:e;V_1:c_1'} + \hat{\sigma}_{\sigma_{D,q_1};S_1:1;V_1:c_1'} \qquad (5)$$

$$D_2 = \hat{\sigma}_{\sigma_{D_1,q_2};S_1:e_2;V_1:c_1''} + \hat{\sigma}_{\sigma_{D,q_3};S_1:e_2;V_1:c_1''} \qquad (6)$$

Now $D_1$ is modified to be formed by deleting rows due to $q_1$ and adding rows by running $q_4$ on $D_2$

$$D_1 = \hat{\sigma}_{\sigma_{D,q};S_1:e_2;V_1:c_1'} + \hat{\sigma}_{\sigma_{D_2,q_4};S_1:e_2;V_1:c_1'} \qquad (7)$$

Thus $D_2$ is dependent on $D_1$ via the query $q_2$ and $D_1$ is dependent on $D_2$ via the $q_4$. If there are rows for $q_2$ and $q_4$ they form a cycle. Fig 2 shows data dependence of equation (6) and (7) graphically. $\Delta D1$, a part of D1, is dependent on $\Delta D2$, a part of D2 and vice versa forming a cycle.

We will only focus on the case where changes in a scenario do not affect any other scenario. This graph, as it grows, can be reduced by keeping the original data cube D immutable and resolving all the nodes of the data graph to D as and when they are made. This effectively means that the queries associated with the scenarios are reduced to contain real dimension values only and not scenario dimension values. This approach makes sure that all the nodes exist independent of other's updates or deletions. This approach also makes sure that no cycles exist in this graph.

### 4.4  Association of a query with a scenario

Let us say a scenario $W_i$ is defined in terms of a set of queries $Q_{w_i} \equiv \{Q_1^i, Q_2^i, \ldots Q_{u_i}^i\}$. Each $Q_j^i$ also contains a set of multiplicative factors $e_j^i$. In general, any of these queries can contain any number of scenario dimension values. In our system, they need to be reduced to real queries. This is done when a query is included in a scenario. At this time $Q_j^i$ is reduced to key queries and list of corresponding real queries stored as their values. The reduced queries also contain reduced factors for measure values. This arrangement is helpful during evaluation of a query (Algorithm 2 and 3). An algorithm to resolve a query containing scenarios into real queries is given below (Algorithm 1). A query Q with a set of multiplicative factors $e_1$, $e_2$ etc. respectively for measures S is to be associated with a scenario $W_i$ whose dimension value is $w_i$.

**ALGORITHM 1.** Association of a query with a scenario

**Input: Date cube D, Scenarios W, a new query Q and the scenario $W_i$ it is to be associated with**
**Output: The new query is included in scenario's queries**

```
if π_Q ≢ Φ
   for each query Q_j^i in η_Q
      Initialize map M = θ_{π_{Q_j^i}}
      Initialize collection C
      for each key (K_j^i)_k in M and value (L_j^i)_k  /*(L_j^i)_k is a collection of queries*/
         if ((K_j^i)_k ∩ Q ∩ <v_1^a, v_2^a...> ≢ Φ ∀ V_a ∈ V, a ≤ n)
            Add (L_j^i)_k to C
         end
      end
      if (C ≢ Φ)
         Q_j^i = Q_j^i ∪ <*,...,w_i, ...*>
         V_j^i = μ_{(L_j^i)_1,(L_j^i)_2...}   /* (L_j^i)_k is present in C*/
         for each query (Q_j^i)_k in V_j^i
            for each S_l in S
               (e_j^i)_l^k = (e_j^i)_l^k × e_l
            end
         end
         Store Q_j^i as key and V_j^i as value
      end
   end
else
   Initialize collection C
   add Q to C
   Query K = Q ∪ <*,..., w_i, ...*>
   Store K as key and C as value
end
```

Where $(L_j^i)_1, (L_j^i)_2 ...$ are elements of collection of C
$(e_j^i)_l^k$ is a multiplying factors for a query $(Q_j^i)_k$ and a measure $S_l$.
/*Text*/ is comment in the algorithm

Extending our example, let us say VP Operations is planning to remove SU2 from panel as he feels SU2 is over-charging. He has a competing offer from a new supplier SU3 who is offering a discount of 10% on SU2's prices and he wants to look how amounts would look like if SU3 was used instead of SU2. He creates a scenario for SU3 $< SU3, \text{Supplier: SU3} >$. Query for scenario SU3 is given below and corresponding rows for scenario SU3 are shown in Table VIII

$$< SU3, \text{Supplier: SU3} > \equiv \hat{\sigma}_{\hat{\sigma}_{D,<2011,SU2,P1,P2>};\text{Volume:1,Cost:0.9;Supplier:SU3}} \tag{8}$$

Table VIII. Data cube for scenario SU3

| Year | Supplier | Product | Volume(MM Tons) | Cost($/Unit) | Amount(MM$) = Cost * Volume |
|------|----------|---------|-----------------|--------------|------------------------------|
| 2011 | SU3      | P1      | 12              | 0.99         | 11.88                        |
| 2011 | SU3      | P2      | 13              | 1.26         | 16.38                        |

VP Operations shares this plan with the sourcing manager and asks him to make a projection for 2012 with SU3 and without SU2. The manager modifies his 2012 scenario to include the changes

$< 2012, \text{Year}: 2012 > \equiv \hat{\sigma}_{\sigma_{D,<2011,SU1,P1,P2>};\text{Volume: 2,Cost:1;Year:2012}} +$

$\hat{\sigma}_{\sigma_{R,<2011,SU3,P1,P2>};\text{Volume: 3,Cost:1;Year:2012}}$ (9)

We need to reduce $\sigma_{R,<2011,SU3,P1,P2>}$ to real queries operating on D.

Query <2011, SU3, P1, P2> can be reduced to real queries by $\mu_{\theta_{\pi_{<2011,SU3,P1,P2>}}}$

$= \mu_{\{<K:2011,SU2,SU3\ P1,P2,V:\{2011,SU2,P1,P2>\text{Volume: 1,Cost: 0.9}\}\}}$
$= \{K: < 2011, SU2, SU3, P1, P2 >, V: \{< 2011, SU2, P1, P2 > \text{Volume: 1, Cost: 0.9}\}\}$ (10)
$\Rightarrow Q_{W_{SU3}} \equiv \{K: < 2011, SU2, SU3, P1, P2 >, V: \{< 2011, SU2, P1, P2 > \text{Volume: 1, Cost: 0.9}\}\}$

Where K is the key and V is the value

$\sigma_{R,<2011,SU3,P1,P2>} = \hat{\sigma}_{\sigma_{D,<2011,SU2,P1,P2>};\text{Volume: 1,Cost:0.9;Supplier:SU3}}$

Hence, $\hat{\sigma}_{\sigma_{R,<2011,SU3,P1,P2>};\text{Volume: 3,Cost:1;Year:2012}}$

$= \hat{\sigma}_{(\hat{\sigma}_{D,<2011,SU2,P1,P2>\text{Volume: 1,Cost:0.9;Supplier:SU3}})\text{Volume: 3,Cost:1;Year:2012}}$

$= \hat{\sigma}_{(\hat{\sigma}_{D,<2011,SU2,P1,P2>;\text{Volume:1,Cost:1;Supplier:SU3}})\text{Volume: 3,Cost:0.9,Year:2012}}$ (11)

$\Rightarrow < 2012, \text{Year}: 2012 > \equiv \hat{\sigma}_{\sigma_{D,<2011,SU1,P1,P2>};\text{Volume: 2,Cost:1;Year:2012}} +$

$\hat{\sigma}_{(\hat{\sigma}_{D,<2011,SU2,P1,P2>;\text{Volume:1,Cost:1;Supplier:SU3}})\text{Volume: 3,Cost:0.9,Year:2012}}$ (12)

Here $Q_{W_{2012}} \equiv$ {K:<2011, 2012, SU3, P1, P2>, V:{<2011, SU2, P1, P2> Volume: 3, Cost: 0.9}},{K: <2011, 2012, SU1, P1, P2> , V: {<2011, SU1, P1, P2> Volume: 3, Cost: 1}}} where K is the key and V is the value. The new plan of the manager for the year 2012 is shown in Table IX

Table IX. Data cube for changed scenario 2012

| Year | Supplier | Product | Volume (MM Tons) | Cost ($/Unit) | Amount($) = Cost * Volume |
|---|---|---|---|---|---|
| 2012 | SU1 | P1 | 20 | 1.0 | 20 |
| 2012 | SU1 | P2 | 22 | 1.5 | 33 |
| 2012 | SU3 | P1 | 36 | 0.99 | 35.64 |
| 2012 | SU3 | P2 | 39 | 1.26 | 49.14 |

### 4.4. Sub-cube derivation

In this section we give an implementation of select modify operator defined in section 3.1 to materialize a sub-cube using a query with our definition and structure of scenario.

**ALGORITHM 2.** Materialize a sub cube

**Input:** Data cube D, Scenarios W, an arbitrary query E for which cube is to be generated
**Output:** A sub-cube

Initialize a Map C which contains scenario dimension value as key and another map M as value. M contains a query as a key and a collection of queries as values.

**for** each scenario $W_i$ in the set of scenarios $\pi_E$
    add $\{w_i: m_i\}$ to C /*$m_i$ is a map*/
    **for** each query $Q_j^i$ in $W_i$
        **if** $(E \cap Q_j^i \cap <v_1^k, v_2^k...> \not\equiv \Phi \; \forall \; V_k \in V)$
            add $\{Q_j^i:\{(Q_j^i)_1, (Q_j^i)_2...\}\}$ to $m_i$
        **end**
    **end**
**end**
**for** each row t in D
    **if** $(E \cap t_d = t_d)$
        accumulator $(t_d, t_{S_l})$
    **end**
    **for** each scenario dimension value $w_i$ in C and value $m_i$
        **for** each occurrence of key $Q_j^i$ in $m_i$
            **if** $(\rho_{Q_j^i \cap E} \cap t_d = t_d)$
                **for** each query $(Q_j^i)_k$ against the key $Q_j^i$ in $m_i$
                      **if** $((Q_j^i)_k \cap t_d = t_d)$
                          Initialize a query $t_R = t_d$
                          **for** each scenario $W_m$ in $\pi_{Q_j^i}$
                              Initialize $V_p$ = dimension of scenario $W_m$
                              $t_R = t_R \setminus <v_1^p, v_2^p, ... v_{u_p}^p> \cup <*,...,w_m,...*>$
                          **end**
                          Initialize $V_p$ = dimension of scenario $W_m$
                          accumulator $(t_R \setminus <v_1^p, v_2^p, ... v_{u_p}^p> \cup <*,...,w_i,...*>, t_{S_l} \times (e_j^i)_l^k)$
                      **end**
                  **end**
            **end**
        **end**
    **end**
**end**

Where

$A \setminus B$ represents relative complement of two queries A and B

Accumulator is a function which takes dimension and measure values of a row and combines them to materialize a row. The accumulator can later be queried to produce the desired sub-cube

Table X shows the rows for the query <2011, 2012, SU1, SU2, SU3, P1, P2>. The first four rows are real and exist in the data cube while the last four rows are hypothetical.

Table X. Materialized sub-cube for an arbitrary query

| Year | Supplier | Product | Volume(MM Tons) | Cost($/Unit) | Amount(MM$) = Cost * Volume |
|---|---|---|---|---|---|
| 2011 | SU1 | P1 | 10 | 1.0 | 10 |
| 2011 | SU1 | P2 | 11 | 1.5 | 16.5 |
| 2011 | SU2 | P1 | 12 | 1.1 | 13.2 |
| 2011 | SU2 | P2 | 13 | 1.4 | 18.2 |
| 2012 | SU1 | P1 | 2 * 10 | 1 * 1.0 | 20 |
| 2012 | SU1 | P2 | 2 * 11 | 1 * 1.5 | 33 |
| 2012 | SU3 | P1 | 3 * 12 | 0.9 * 11 | 35.64 |
| 2012 | SU3 | P2 | 3 * 13 | 0.9 * 14 | 49.14 |

### 4.5. Aggregating measure values without row materialization

If E is an arbitrary query, its measure values can be accumulated (using a function like sum, average etc.) without materializing the scenario rows. Although this can also be achieved by using an algorithm similar to algorithm 1, we present an alternate to avoid η operator which makes combinations of scenario dimension values in a query q.

---
**ALGORITHM 3.** Query evaluation on the system (real and hypothetical taken together)

---
**Input: Date cube D, Scenarios W, and a query E to evaluate**
**Output: Accumulated output of measures in qualified rows**
Initialize Map of queries C with key as a query and values as a list of queries
**for** each scenario $W_i$ in the set of scenarios $\pi_E$
    **for** each query $Q_j^i$ in $W_i$
        Query $q_j^i = \rho_{Q_j^i \cap E}$
        **if** ($q_j^i \cap <v_1^k, v_2^k...> \not\equiv \Phi \; \forall \; V_k \in V$)
            add $\{q_j^i: \{(Q_j^i)_1, (Q_j^i)_2...\}\}$ to C
        **end**
    **end**
**end**
**for** each row t in D
    **if** (E ∩ $t_d = t_d$)
        accumulator ($t_{S_1}, t_{S_2}, ...$)
    **end**
    **for** each key query $q_j^i$ in C
        **if** ($q_j^i \cap t_d = t_d$)
            **for** each query $(Q_j^i)_k$ against the key $q_j^i$ in C
                **if**($(Q_j^i)_k \cap t_d = t_d$)
                      accumulator ($t_{S_1} \times (e_j^i)_1^k, t_{S_2} \times (e_j^i)_2^k ...$)
                **end**
            **end**
        **end**
    **end**
**end**

Where
accumulator is a function which takes measures of eligible rows and applies an operation like sum, average etc. on them. It can later be queried for results later.

In our running example, manager submits a report comparing the total amount used in 2011 using query q1<2011, SU1, SU2, P1, P2> as shown in Table I and required in 2012 using query q2<2012, SU1, SU3, P1, P2>, as shown in Table X

Sum of amount ($\sigma_{R,<2011,SU1,SU2,P1,P2>}$) = MM$ (10 + 16.5 + 13.2 + 18.2) = MM $57.9

Sum of amount ($\sigma_{R,<2012,SU1,SU3,P1,P2>}$) = MM$137.78

### 4.5 Choice of data structure for scenario queries

When an arbitrary query is associated with a scenario, it is reduced into real queries. These real queries are stored against a key. For evaluation of another arbitrary query (called evaluation query below), we need some information preserved. The operators help generate this information and the data structures facilitate its storage and use during evaluation of queries. The following information is conserved:

1. Scenario dimension value in the original query with which it is associated. In our running example, if a query Q<2011, SU1, P1> is associated with a scenario P3 and its key query is stored without scenario dimension value P3 i.e. K<2011, SU1, P1> then during evaluation of an evaluation query E<2011, SU1, P3>, E ∩ K (which appears in the Algorithm 3) returns no dimension values for product and is wrongly excluded from accumulation. This is addressed by keeping the scenario dimension value in scenario's keys i.e. K should be <2011, SU1, P1, P3>

2. The key query. In our running example let us associate the query Q <2011, SU3, P1, P2> with a new product scenario P3. Since SU3 is a scenario dimension value, Q is reduced to a real query Q' <2011, SU2, P1, P2>. Now, if evaluation query is E<2011, SU3, P3> and Q is not stored as the key K<2011, SU3, P1, P2, P3>, Q' alone doesn't suffice to discover if SU3 was present in Q. Thus, it is important to store K along with Q'

3. The mapping between the key query and the derived real queries e.g. in the above example, if another query Q1<2012, SU3, P1, P2> is also associated with P3, it will be reduced to real query Q1' <2011, SU2, P1, P2> and stored against key query K1<2012, SU3, P1, P2, P3>. Now, if evaluation query is E <2011, SU3, P3> and the mapping from key queries to derived real queries is not maintained, there is no way to discover which of the derived queries Q' and Q1' should be picked. Thus, it is important to have a mapping from key query to list of real queries as value. If such a relationship exists, E ∩ K and E ∩ K1 can be tested to contain dimension values for each dimension. Derived queries of only those keys for which the condition holds will be considered. In this case Q is such a query

4. Key query in atomic form. If the original query contains at least two dimension values for a dimension and at least one of them is a scenario dimension value, it is not possible to identify the right derived queries while running an evaluation query e.g. a query Q <2011, SU2, SU3, P1, P2> is associated with a product scenario P3. It generates real queries Q1'<2011, SU2, P1, P2> due to SU2 and Q2' <2011, SU2, P1, P2> due to SU3 against the key query K<2011, SU2, SU3, P1, P2, P3>. If evaluation query is E<2011, SU3, P3>, according to Algorithm 3, E ∩ K = <2011, SU3, P3> contains dimension values for all dimensions and $\rho_{E \cap K}$ = <2011, *, *> makes both Q1' and Q2' valid for accumulation. This is incorrect since SU2 is absent in E; rows of SU2 should have been excluded. Thus, only Q2' should have been selected. Similarly, a query containing multiple scenario dimension value will also face the same issues.
   4.1. This can be corrected if Q is divided into atomic queries viz. $Q_1$<2011, SU2, P1, P2> and $Q_2$<2011, SU3, P1, P2>. Then, for $K_1$<2011, SU2, P1, P2, P3> as key, Q1' is value and for $K_2$<2011, SU3, P1, P2, P3> as key, Q2' is value. E ∩ $K_1$ in this case will not contain dimension values for Supplier and hence will not be considered. E ∩ $K_2$ will be considered and will give Q2' for accumulation. We use operator η to generate atomic queries. Each atomic query is used as a key and the generated real queries, as a result of operator μ, are stored as value.
   4.2. Division into atomic queries is not required when only real dimension values are present in the original query e.g. if a query Q1<2011, SU1, SU2, P1, P2> is associated with a new scenario P3, it will be stored as K1<2011, SU1, SU2, P1, P2, P3> and supplier dimension values are not divided into <2011, SU1, P1, P2, P3> and <2011, SU2, P1, P2, P3>. This is because the real dimension values appear in the dimension values of real rows and it is possible to ascertain which real rows correspond to which dimension values. If the evaluation query is E<2011, SU1, SU3, P1, P2, P3>, E ∩ K1 contains all the dimensions and $\rho_{E \cap K1}$ = <2011, SU1, P1, P2, P3> will make sure that only SU1 rows are included for accumulation.
   4.3. Scenarios are defined in terms of arbitrary queries. For every dimension in an arbitrary query, dimension values can be present in one of the eight ways as shown in Table XI

Table XI. Various ways dimension values can be arranged

|  | No Real | Single Real | Multiple Real |
|---|---|---|---|
| No Scenario | - | √ | √ |
| Single Scenario | √ | × | × |
| Multiple Scenario | × | × | × |

Cell 'No Scenario' – 'Single Real' means a single real dimension value is present for a dimension. Cell 'No Scenario' – 'Multiple Real' means multiple real dimension values are present. Cell 'Single Scenario'- 'Single Real' means a single dimension value is present for both real and scenario dimension values and so on. Cell 'No Scenario' – 'No Real' is ignored as this query returns empty set. Dividing an arbitrary query into an atomic query reduces five cases which are crossed out. This includes all the cases which may have multiple scenario dimension values because of the observation made in 4.1. Multiple real dimension values can be handled as explained in 4.2. A key query, containing a single real or scenario dimension value for a dimension never faces any issue as the queries in its value are known to be associated with the single dimension value. We have provided a single solution for all the remaining cases. It is important to do so as a query contains dimension value for all the dimensions and distribution of dimension values of each dimension may fall in any of the above cases.

## 5. SUMMARY

In this paper, we have described a way to introduce new dimension values for what-if analysis without changing the read-only data cube. To achieve this, we have extended data graph model [Balmin et. al., 2000] and used query system described by [Li et. al., 2004]. As a result we can evaluate a query on the whole data cube, both real and hypothetical simultaneously. In our approach, new rows are not physically added to the data cube but are stored as rules as shown in Algorithm 1. These rules can help simulate hypothetical rows as shown in Algorithm 2. Analytical systems can use Algorithm 3 for fast, ad-hoc query evaluation and aggregation.

## 6. FUTURE WORK

We need to obtain performance metrics and compare them with the industry standards. Other problems that can be solved by extending this approach are what-if analysis on calculated measures and deriving scenario rules by taking the scenario rows from an external source. Our approach in this paper keeps scenarios in global scope, however, it can also be extended to keep scenarios private for each user as well.